\documentclass{sigchi}

\usepackage{multirow}
\usepackage{amsmath}
\usepackage{graphicx,subfigure}

\newcommand\given[1][]{\:#1\vert\:}
\newcommand{\R}{\mathbb{R}}

\CopyrightYear{2020}
\setcopyright{acmlicensed}
\doi{https://doi.org/XX.XXXX/XXXXXXX.XXXXXXX}
\isbn{XXX-X-XXXX-XXXX-X/20/XX}

\usepackage{balance}       
\usepackage{graphics}      
\usepackage[T1]{fontenc}   
\usepackage{txfonts}
\usepackage{mathptmx}
\usepackage[pdflang={en-US},pdftex]{hyperref}
\usepackage{color}
\usepackage{booktabs}
\usepackage{textcomp}

\usepackage{microtype}        
\usepackage{ccicons}          

\usepackage{todonotes}

\def\plaintitle{A Survey on Interactive Reinforcement Learning: Design Principles and Open Challenges}

\def\emptyauthor{}
\def\plainkeywords{Interactive Machine Learning; Interactive Reinforcement Learning; Human-agent Interaction.}

\makeatletter
\def\url@leostyle{%
  \@ifundefined{selectfont}{
    \def\UrlFont{\sf}
  }{
    \def\UrlFont{\small\bf\ttfamily}
  }}
\makeatother
\def\pprw{8.5in}
\def\pprh{11in}

\definecolor{linkColor}{RGB}{6,125,233}
\hypersetup{%
  pdftitle={\plaintitle},
  pdfauthor={\emptyauthor},
  pdfkeywords={\plainkeywords},
  pdfdisplaydoctitle=true, 
  bookmarksnumbered,
  pdfstartview={FitH},
  colorlinks,
  citecolor=black,
  filecolor=black,
  linkcolor=black,
  urlcolor=linkColor,
  breaklinks=true,
  hypertexnames=false
}

\begin{document}

\title{\plaintitle}

\numberofauthors{3}
\author{%
  \alignauthor{Christian Arzate Cruz\\
    \affaddr{The University of Tokyo}\\
    \affaddr{Tokyo, Japan}\\
    \email{arzate.christian@gmail.com}}\\
  \alignauthor{Takeo Igarashi\\
    \affaddr{The University of Tokyo}\\
    \affaddr{Tokyo, Japan}\\
    \email{takeo@acm.org}}\\
}

\maketitle

\begin{abstract}
  Interactive reinforcement learning (RL) has been successfully used in various applications in different fields, which has also motivated HCI researchers to contribute in this area. In this paper, we survey interactive RL to empower human-computer interaction (HCI) researchers with the technical background in RL needed to design new interaction techniques and propose new applications. We elucidate the roles played by HCI researchers in interactive RL, identifying ideas and promising research directions. Furthermore, we propose generic design principles that will provide researchers with a guide to effectively implement interactive RL applications.
\end{abstract}




\section{Introduction}
Machine learning (ML) is making rapid advances in many different areas. However, standard ML methods that learn models automatically from training data sets are insufficient for applications that could have a great impact on our everyday life, such as autonomous companion robots or self-driving cars. Generally speaking, to enable their use in more complex and impactful environments, ML algorithms have to overcome further obstacles: how to correctly specify the problem, improve sampling efficiency, and adapt to the particular needs of individual users. 

Interactive ML (IML) proposes to incorporate a human-in-the-loop to tackle the aforementioned problems in current ML algorithms \cite{Fails2003, Holzinger2016, Holzinger2016_Ants}. This configuration intends to integrate human knowledge that improves and/or personalizes the behavior of agents. Reinforcement learning (RL) is one ML paradigm that can benefits from an interactive setting \cite{Suay2011}. Interactive RL has been successfully applied to a wide range of problems in different research areas \cite{Arakawa2018, Warnell2018, Krening2018}. However, interactive RL has not been widely adopted by the human-computer interaction (HCI) community. 

In the last few years, the two main roles that HCI researchers have played in IML are (1) designing new interaction techniques \cite{Simard2017} and (2) proposing new applications \cite{Yang2017, Chang2017}. To enable HCI researchers to play those same roles in interactive RL, they need to have adequate technical literacy. In this paper, we present a survey on interactive RL to empower HCI researchers and designers with the technical background needed to ideate novel applications and paradigms for interactive RL.

To help researchers perform role (1) by describing how different types of feedback can leverage an RL model for different purposes. For instance, the user can personalize the output of an RL algorithm based on the user's preferences \cite{Abel2017}\cite{Leike2018}, facilitate learning in complex tasks \cite{Knox2009}, or guide exploration to promising states in the environment \cite{Garcia2015}. 

Finally, we help researchers perform role (2) by proposing generic design principles that will provide a guide to effectively implement interactive RL applications. We pay special attention to common problems in multiple research fields that constrain the performance of interactive RL approaches; solving these problems could make interactive RL applications effective in real-world environments.

\subsection{Scope of this Paper}
The literature on RL that could potentially be applied to interactive RL is extensive. For that reason, we focus on works that have been successfully applied to RL methods that use a human-in-the-loop. We further narrow the reach of this survey by considering only works from three research areas, robotics, HCI, and Game AI, published between $2010$ and $2019$. Among these papers, we picked those that covered all the main research directions in the three research areas. We chose these three fields because they use interactive RL systems in similar ways, enabling us to propose generic design rules for interactive RL that will be useful for a wider audience. 

Some surveys have covered larger topics, such as IML in general \cite{Amershi2014} or IML for health informatics \cite{Holzinger2016}. Other surveys have focused on particular design dimensions, such as types of user feedback \cite{Leike2018} or the evaluation of results \cite{Anjomshoae2019}. Surveys have also been done on particular subjects within the interactive RL research area, such as safe RL \cite{Garcia2015}, the agent alignment problem \cite{Leike2018}, and inverse RL \cite{Zhifei2012}. In contrast, our survey focuses only on approaches that can be applied to an interactive RL setting. 

These constraints allow us to construct a detailed analysis for HCI researchers. Additionally, our survey methodology is favorable for constructing design principles for interactive RL that are generic enough to be applied in physical or simulated environments.

\section{Reinforcement Learning (RL)}
\label{sec::introRL}
The \textit{reinforcement learning} (RL) paradigm is based on the idea of an agent that learns by interacting with its environment \cite{Sutton1996, Kaelbling1996}. The learning process is achieved by an exchange of signals between the agent and its environment; the agent can perform actions that affect the environment, while the environment informs the agent about the effects of its actions. Additionally, it is assumed that the agent has at least one goal to pursue and -- by observing how its interactions affect its dynamic environment -- it learns how to behave to achieve its goal.

Arguably, the most common approach to model RL agents is the \textit{Markov decision process} (MDP) formalism. An MDP is an optimization model for an agent acting in a stochastic environment \cite{Puterman2014}; that is defined by the tuple $\langle S,A,T,R,\gamma\rangle$ , where $S$ is a set of states; $A$ is a set of actions; $T \colon S \times A \times S \rightarrow [0, 1]$ is the \textit{transition function} that assigns the probability of reaching state $s'$ when executing action $a$ in state $s$, that is, $T(s' \given s,a) = P(s' \given a,s)$; $R \colon S \times A \rightarrow \R$ is the \textit{reward function}, with $R(s, a)$ denoting the immediate numeric reward value obtained when the agent performs action $a$ in state $s$; and $\gamma \in [0, 1]$ is the discount factor that defines the preference of the agent to seek immediate or more distant rewards.

The reward function defines the goals of the agent in the environment; while the transition function captures the effect of the agent's actions for each particular state. 

\section{Interactive Reinforcement Learning}
\label{sec::IntroIML}
In a typical RL setting, the RL designer builds a reward function and chooses/designs an RL method for the agent to learn an optimal policy. For a survey of the literature on RL, we refer the reader to \cite{Arulkumaran2017, Polydoros2017, Ghavamzadeh2015}.  

In contrast, an interactive RL approach involves a human-in-the-loop that tailors specific elements of the underlying RL algorithm to improve its performance or produce an appropriate policy for a particular task. One of the first approaches that incorporated a human-in-the-loop in an RL algorithm is the Cobot software agent \cite{Isbell2001}. This interactive method implements a crowd-sourced environment where multiple users work together to create the reward function for a virtual agent. However, a formal introduction to the term interactive RL was later proposed in \cite{Thomaz2005}. 

\begin{figure}[!htb]
\begin{center}
\includegraphics[width = 1.0 \columnwidth ]{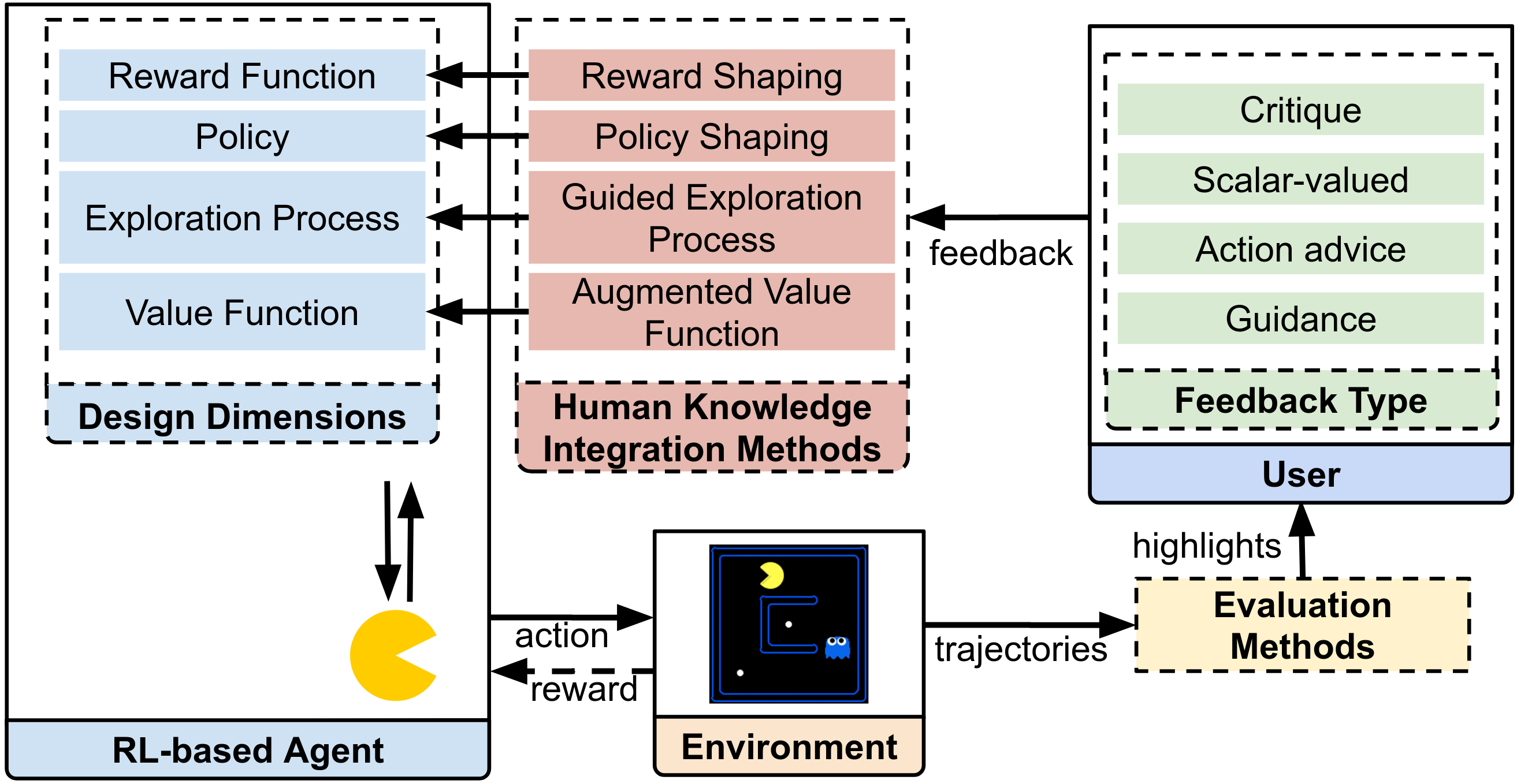}
		\caption{The interactive RL architecture.}
\label{Fig:iRL}
\end{center}
\end{figure}

Figure \ref{Fig:iRL} presents a generic interactive RL architecture and shows the interaction channels that allow the user to guide the low-level components of an RL algorithm. The foundation of an interactive RL approach is an RL-based agent that learns by interacting with its environment according to a specific RL algorithm. Next, the user observes and evaluates the resulting agent behavior and gives the agent feedback. Human feedback can be delivered through multiple human knowledge integration methods to a specific element (design dimension) of the underlying RL algorithm. 

The interactive RL paradigm has the advantage of integrating prior knowledge about the task the RL-based agent is trying to solve. This concept has proven to be more effective than automatic RL algorithms \cite{Fails2003, Knox2012_2}. Moreover, human feedback includes prior knowledge about the world that can significantly improve learning rates \cite{Dubey2018}. Arguably, any RL algorithm will more quickly find a high-earning policy if it includes human knowledge. However, it is still unclear how to design efficient interactive RL methods that generalize well and can be adapted to feedback from different user types. 

In the rest of this paper, we will detail all the elements of interactive RL that we have introduced so far.

\section{Interactive RL Testbeds}
\label{sec::Testbeds}
\begin{figure*}[t!]
\centering     
\subfigure[ ]{\label{fig:a}\includegraphics[height=28mm]{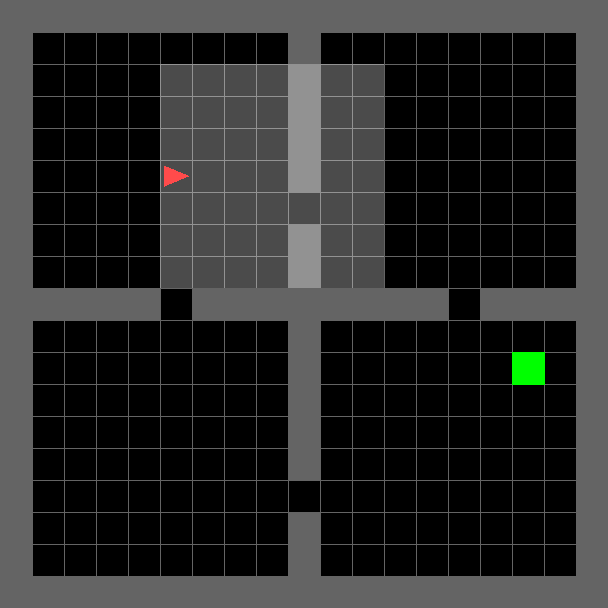}}
\subfigure[ ]{\label{fig:b}\includegraphics[height=28mm]{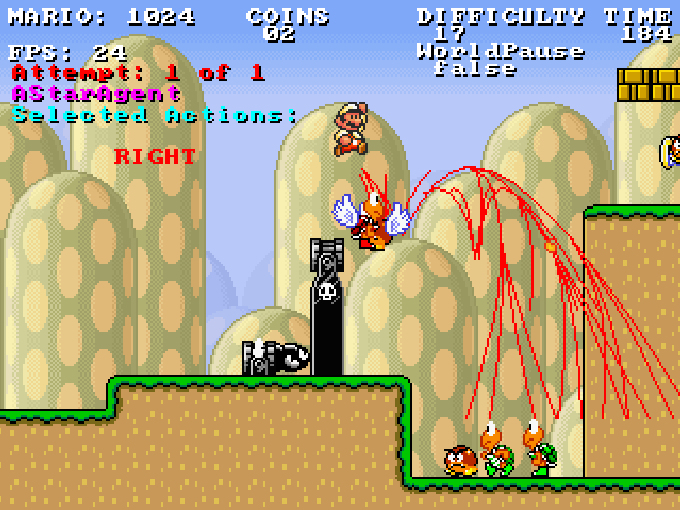}}
\subfigure[ ]{\label{fig:b}\includegraphics[height=28mm]{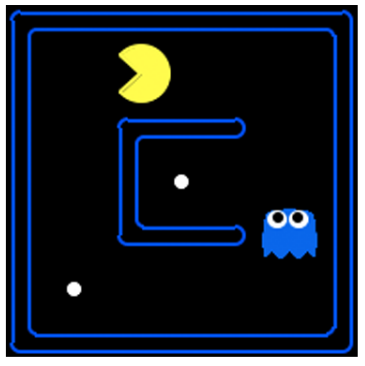}}
\subfigure[ ]{\label{fig:b}\includegraphics[height=28mm]{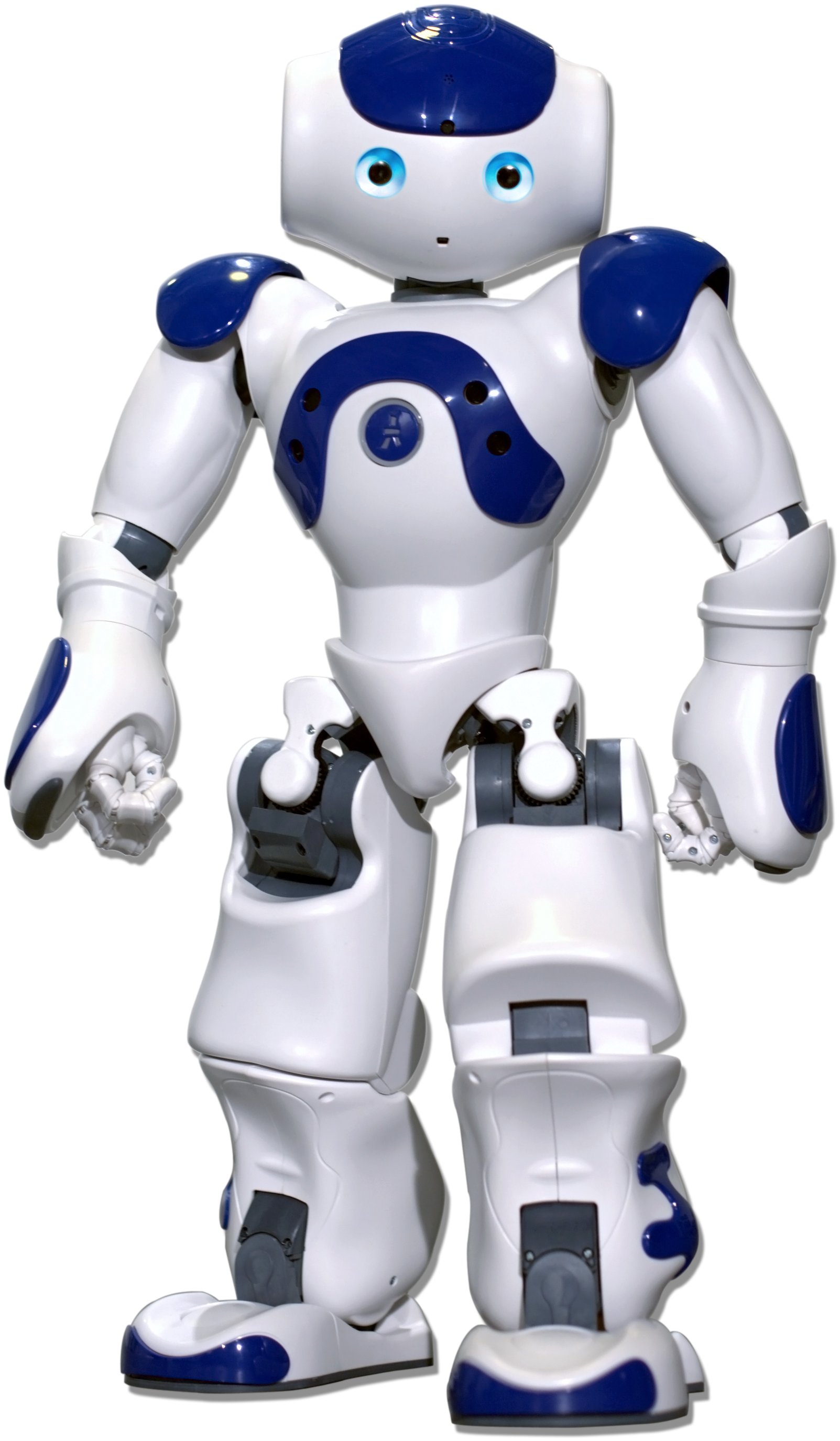}}
\subfigure[ ]{\label{fig:b}\includegraphics[height=28mm]{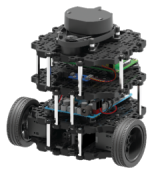}}
\subfigure[ ]{\label{fig:b}\includegraphics[height=28mm]{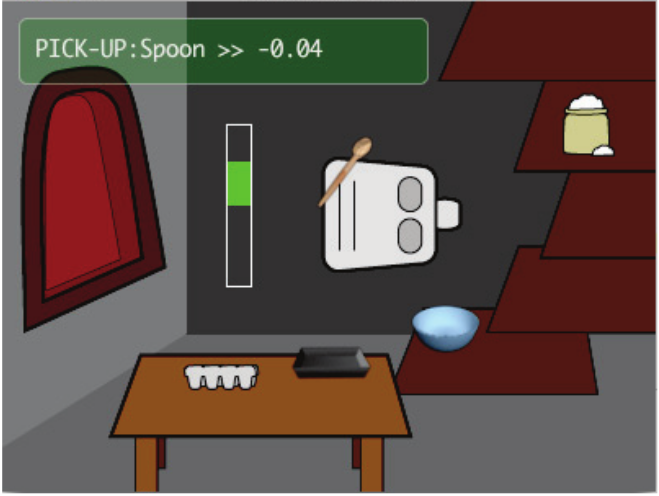}}

\caption{Selected testbeds for interactive RL. \scriptsize (a) GridWorld. (b) Mario AI benchmark. (c) Pac-Man. (d) Nao Robot. (e) TurtleBot. (f) Sophie's kitchen.}
\label{fig::testbeds}
\end{figure*}

In Figure \ref{fig::testbeds}, we present examples of the most common testbeds in interactive RL research. The \textit{GridWorld} platform is a classic environment for testing RL algorithms. The small state-space size of this testbed allows for fast experiments, and its simplicity does not demand specialized knowledge from the users. These characteristics make GridWorld popular among researchers in the human-computer interaction (HCI) field \cite{Krening2018, Krening2018_2, Krening2019}. The \textit{Mario AI benchmark} \cite{Karakovskiy2012} is a clone of Nintendo's platform game named \textit{Super Mario Bros.} This game is a testbed with a large state-space that offers the conditions needed to test interactive RL algorithms aimed to personalize bot behaviors and create fast evaluation techniques. Similarly, we have the \textit{Pac-Man} game that can reduce the size of its state-space to make it more manageable. The \textit{Nao robot} and the \textit{TurtleBot} are popular platforms in robotics that are perfect for testing natural ways to interact with human users. The \textit{Sophie's kitchen} \cite{Thomaz2006} platform is designed as an online tool for interactive RL that explores the impact of demonstrating uncertainty to the users. Finally, the main focus of the OpenAI Gym platform \cite{OpenAI2016} is designing and comparing RL algorithms.

Generally speaking, these are the classifications of testbeds for each research area included in our survey:       

\begin{itemize}
\item \textbf{Robotics:} GridWorld environments (physical and virtual), OpenAI Gym, soccer simulation, Nao robot, Nexi robot, TurtleBot, shopping assistant simulator.

\item \textbf{Game AI:} Pac-Man, Mario AI benchmark, Street Fighter game, Atari emulator.

\item \textbf{HCI:} Pac-Man with small state-space, GridWorld environments.
\end{itemize}

\subsection{Applications}
We can find applications for interactive RL, such as teaching robots how to dance \cite{Meng2014}, creating adaptable task-oriented dialogue systems \cite{Shah2016}, and learning the generation of dialogue \cite{Li2016deep}. Furthermore, there is the opportunity to adapt current automatic RL applications to an interactive setting, such as procedural content generation for video games \cite{Khalifa2020}, generating music \cite{Jaques2016}, and modeling social influences in bots \cite{Jaques2018social}.

\section{Design Guides for interactive RL}
This section provides HCI researchers with the foundations to design novel interaction techniques for interactive RL.

\subsection{Design Dimensions}
\label{sec::DesignDimensions}
\begin{table*}[!ht]
\scriptsize 
\begin{center}
\begin{tabular}{l l l l l l l}
\toprule
\textbf{Design Dimension} & \textbf{Testbed} & \textbf{Interaction}  & \textbf{Initiative} & \textbf{HKI} & \textbf{Feedback} & \textbf{Algorithms}\\
\midrule
Reward Function & Robot in maze-like environment \cite{Arakawa2018}  & FE & Passive & RS using HF $+$ ER & Critique & DQL \cite{Lin1992, Mnih2013}\\
			  ~ & Navigation simulation \cite{Arumugam2018} & GUI & Passive & Advantage Function & Critique & DAC\\
			  ~ & Sophie's Kitchen game \cite{Thomaz2005, Thomaz2006} & GUI & Passive, Active & RS using HF $+$ EF & Critique & QL \cite{Watkins1992}, HRL \\
			  ~ & Bowling game \cite{Warnell2018} & GUI  & Passive & RS $+$ HF & Scalar-valued & DQL \\
			  ~ & Shopping assistant, GridWorld \cite{Mindermann2018} & GUI & Active & Active IRD & Queries & Model-based RL \\
			  ~ & Mario, GridWorld, Soccer simulation  \cite{Rosenfeld2018} & Coding & Passive & PBRS & Heuristic Function & QL, QSL, QL($\lambda$), QSL($\lambda$) \\
			  ~ & Navigation simulation \cite{Tenorio2010} & VC & Passive & RS using HF $+$ ER & AcAd & SARSA \cite{Rummery1994}, SARSA($\lambda$) \\
			  ~ & Atari, robotics simulation \cite{Christiano2017} & GUI & Active & RS using HF & Queries & DRL \\

\midrule
Policy & GridWorld, TurtleBot robot \cite{Macglashan2017} & GUI, GC & Passive & PS & AcAd & AC($\lambda$) \cite{Bhatnagar2009, Singh1996} \\	
	 ~ & GridWorld \cite{Krening2018} & VC & Passive & PS & Critique, AcAd & BQL \cite{Dearden1998} \\
	 ~ & Pac-Man, Frogger \cite{Griffith2013} & GUI & Passive & PS & Critique & BQL \\

\midrule
Exploration Process & Pac-Man, Cart-Pole simulation \cite{Yu2018} & GUI & Passive & GEP & AcAd & QL \\
				  ~ & Simulated cleaning Robot \cite{Cruz2015, Cruz2016} & VC & Passive & GEP & AcAd & SARSA\\
				  ~ & Pac-Man \cite{Amir2016} & GUI & Active & GEP & AcAd  & SARSA($\lambda$)\\
				  ~ & Pac-Man \cite{Fachantidis2017} & GUI & Active & Myopic Agent & AcAd & QL, QRL\\	
				  ~ & Sophie's Kitchen game \cite{Thomaz2006} & GUI & Active & ACTG & Guidence & QL \\ 
				  ~ & Street Fighter game \cite{Arzate2018} & Not apply & Passive & EB using Safe RL & Demonstration & HRL \\
				  ~ & Nao Robot \cite{Suay2011} & GUI & Passive & ACTG & Guidence & QL \\
			  ~ & Nexi robot  \cite{Knox2013} & AT $+$ CT & Passive & Myopic Agent & AcAd & SARSA($\lambda$) \\
\midrule
Value Function & Mountain Car simulation \cite{Knox2010} & GUI & Passive & Weighted VF & Demonstration & SARSA($\lambda$) \\
			 ~ & Keepaway simulation \cite{Taylor2011} & GUI & Passive &  Weighted VF & Demonstration & SARSA \\
			 ~ & Mario, Cart Pole \cite{Brys2015} & Not apply & Passive & Initialization of VF & Demonstration & QL($\lambda$) \\

\bottomrule
\end{tabular}
\end{center}
\caption{Selected works for each design dimension. \scriptsize HKI=human knowledge integration, RS=reward shaping, PS=policy shaping, AcAd=action advice, VF=value function, GUI=grafical user interface, FE=facial expression, VC=voice command, GC=game controller, AT=artifact, CT=controller, HF=human feedback, ER=environmental reward, GEP=guided exploration process, PBRS=potential based reward shaping, ACTG=actions containing the target goal, QL=q-learning, $\lambda$= with elegibility traces, DQL= deep q-learning, QRL= R-learning, QSL=QS-learning, AC($\lambda$)=actor-critic with elegibility traces, DAC= deep actor-critic, DRL= deep reinforcement learning, BQL=bayesian q-learning, HRL=hierarchical reinforcement learning, SARSA=state-action-reward-state-action, IRD= inverse reward design, Mario=Super Mario Bros.}
\label{table:DesignDimensions}
\end{table*}

In this section, we analyze how human feedback is used to tailor the low-level modules of diverse RL algorithms; we call this form of interaction between users and RL algorithms \textit{design dimensions}. Furthermore, the summary of selected works in Table \ref{table:DesignDimensions} presents a concise and precise classification that helps to compare different design dimensions.

\subsubsection{Reward Function}
\label{subsec::RewardShaping}
All works classified in this design dimension tailor the reward function of the RL-based algorithm using human feedback. The main objectives of this feedback are to speed up learning, customize the behavior of the agent to fit the user's intentions, or teach the agent new skills or behaviors.

Designing a reward function by hand is trivial for simple problems where it is enough to assign rewards as $1$ for winning and $0$ otherwise. For example, in the simplest setup of a GridWorld environment, the agent receives $-1$ for each time step and $1$ if it reaches the goal state. However, for most problems that an agent can face in a real-world environment, the selection of an appropriate reward function is key — the desired behaviors have to be encoded in the reward function.   

The reward design problem is therefore challenging because the RL designer has to define the agent's behaviors as goals that are explicitly represented as rewards. This approach can cause difficulties in complex applications where the designer has to foresee every obstacle the agent could possibly encounter. Furthermore, the reward function has to be designed to handle trade-offs between goals. These reward design challenges make the process an iterative task: RL designers alternate between evaluating the reward function and optimizing it until they find it acceptable. This alternating process is called \textit{reward shaping} \cite{Ng1999}. Ng et al. present a formal definition of this concept in \cite{Ng1999}. 

Reward shaping (RS) is a popular method of guiding RL algorithms by human feedback \cite{Ng1999, Wiewiora2003}. In what is arguably the most common version of RS, the user adds extra rewards that enhance the environmental reward function \cite{Tenorio2010, Arakawa2018, Knox2012_2} as $R' = R + F$, where $F \colon S \times A \times S \rightarrow \R$ is the \textit{shaping reward function} \cite{Ng1999}. A hand-engineered reward function has been demonstrated to increase learning rates \cite{Ng1999, Tenorio2010, Arakawa2018, Rosenfeld2018, Dodson2011, Elizalde2008, Elizalde2012, Korpan2017, Wang2018, McGregor2017, Griffith2013, Li2013, Amir2016, Fachantidis2017}. In contrast, some approaches use human feedback as the only source of reward for the agent \cite{Warnell2018, Knox2013, Arumugam2018}. 

Using RS is also beneficial in sparse reward environments: the user can provide the agent with useful information in states where there are no reward signals from the environment or in  highly stochastic environments \cite{Wiewiora2003}. Another advantage of RS is that it gives researchers a tool with which to better specify, in a granular manner, the goals in the current environment. That is, the computational problem is specified via a reward function \cite{SuttonRHypothesis}.

On the other hand, we need to consider the \textit{credit assignment problem} \cite{Sutton1985temporal, Agogino2004unifying}. This difficulty emerges from the fact that, when a human provides feedback, it is applied to actions that happened sometime in the past — there is always a delay between the action’s occurrence and the human response. Furthermore, the human's feedback might refer to one specific state or an entire sequence of states the bot visited. 

\textit{Reward hacking} is another negative side effect resulting from a deficient reward design \cite{Knox2012_3, Ho2015, Amodei2016, Russell2016}. This side effect can cause nonoptimal behaviors that fail to achieve the goals or intentions of the designers. Usually, these kinds of failure behaviors arise from reward functions that do not anticipate all trade-offs between goals. For example, Clark et al. presented an agent that drives a boat in a race \cite{Dario2016}. Instead of moving forward to reach the goal, the agent learned that the policy with the highest reward was to hit special targets along the track.

\subsubsection{Policy}
In the policy design dimension, we include interactive RL methods that augment the policy of an agent using human knowledge. This process is called \textit{policy shaping} (PS) \cite{Griffith13}.

Some authors also classify approaches that use a binary critique as feedback \cite{Griffith13}. However, when an action is labeled as ``bad'' in this type of algorithm, it is generally pruned from the action set at the corresponding action-state, which creates a bias in the action selection of the agent, not its policy. Consequently, we categorize this method as a guided exploration process. 

The PS approach consists of formulating human feedback as action advice that directly updates the agent’s behavior. The user can interact with the RL algorithm using an action advice type of human feedback. For this design dimension, we therefore need access to the (partial) policy of an (expert) user in the task the agent is learning to perform. 

One advantage of the PS approach is that it does not rely on the representation of the problem using a reward function. Therefore, in real-life scenarios with many conflicting objectives, the PS approach might make it easier for the agent to indicate if its policy is correct, rather than trying to explain it through a reward-shaping method. Nevertheless, the user should know a near-optimal policy to improve the agent’s learning process. Even though the effect of the feedback’s quality has been investigated in interactive RL \cite{Fachantidis2017}, more research on this topic is needed to better understand which design dimension is least sensitive to the quality of feedback.

\subsubsection{Exploration Process}
RL is an ML paradigm based on learning by interaction \cite{Sutton2011}. To learn a given task, the agent aims to maximize the accumulated reward signal from the environment. This is achieved by performing the actions that have been effective in the past — a learning step called \textit{exploitation}. However, to discover the best actions to achieve its goal, the agent needs to perform actions that have not been previously tried — a learning step called \textit{exploration}. Therefore, to learn an effective policy, the agent has to compromise between exploration and exploitation. 

The sampling inefficiency of basic RL methods hinders their use in more practical applications, as these algorithms could require millions of samples to find an optimal policy \cite{Christiano2017, Mnih2015}. An approach called the \textit{guided exploration process} can mitigate this problem. 

The guided exploration process aims to minimize the learning procedure by injecting human knowledge that guides the agent’s exploration process to states with a high reward \cite{Yu2018, Cruz2015, Amir2016, Arzate2018}. This method biases the exploration process such that the agent avoids trying actions that do not correspond to an optimal policy. This design dimension also assumes that the user understands the task well enough to identify at least one near-optimal policy. Despite this assumption, there is empirical evidence indicating that using human feedback (i.e., guidance) to direct the exploration process is the most natural and sample-efficient interactive RL approach \cite{Suay2011}. 

For example, in \cite{Suay2011, Thomaz2006}, users direct the agent's attention by pointing out an object on the screen. Exploration is driven by performing actions that lead the agent to the suggested object. This interaction procedure thus requires access to a model of the environment, so the agent can determine what actions to perform to get to the suggested object. 

Another popular approach to guiding the exploration process consists of creating myopic agents \cite{Knox2009, Knox2013, Amir2016, Fachantidis2017}. This kind of shortsighted agent constructs its policy by choosing at every time-step to perform the action that maximizes the immediate reward, according to the action suggested by the user; the action’s long-term effects are not considered. Although this guiding approach creates agents that tend to overfit, it has been successfully used in different applications \cite{Knox2009, Knox2013, Amir2016, Fachantidis2017}.  

\subsubsection{Value Function}
A \textit{reward function} defines the objectives of the agent as immediate rewards that map state or state-action pairs to a scalar value, while a \textit{value function} is an estimate of the expected future reward from each state when following the current policy. Computing a value function is a necessary intermediate step to find a \textit{policy} \cite{SuttonVFHypothesis}. The main approach of the value function design dimension is creating an \textit{augmented value function}.

The procedure to augment a value function consists of combining a value function created by human feedback with the value function of the agent. This procedure has been demonstrated to bias the behavior of the agent and accelerate the learning rate \cite{Brys2015, Knox2009, Taylor2011}. There are too few studies on this design dimension to conclusively compare its performance to other design dimensions \cite{Knox2010}. Nonetheless, the reuse of value functions might be a good way to minimize the human feedback required; an expert's value function could be used in multiple scenarios that share similar state-action spaces.

\subsubsection{Design Dimension Alternatives}
So far, we have explained how RL experts can inject human knowledge through the main components of a basic RL algorithm. There are other ways, however, to interact with low-level features of particular RL approaches. Next, we will explain the two main design dimension alternatives for an interactive RL setting.

\textit{Function Approximation} (FA) allows us to estimate continuous state spaces rather than tabular approaches such as tabular Q-learning in a GridWorld domain. For any complex domain with a continuous state space (especially for robotics), we need to represent the value function continuously rather than in discrete tables; this is a scalability issue. The idea behind FA is that RL engineers can identify patterns in the state-space; that is, the RL expert is capable of designing a function that can measure the similarity between different states.

FA presents an alternative design dimension that can work together with any other type of interaction channel, which means the implementation of an FA in an interactive RL algorithm does not obstruct interaction with other design dimensions. For example, Warnell et al. proposed a reward-shaping algorithm that also uses an FA to enable their method to use raw pixels as input from the videogame they used as a testbed \cite{Warnell2018}. However, their FA is created automatically.

As far as we know, the paper in \cite{Rosenfeld2018} is the only work that has performed user-experience experiments for an interactive RL that uses hand-engineered FAs to accelerate the base RL algorithm. Rosenfeld et al. asked the participants of the experiment to program an FA for a soccer simulation environment. The FAs proposed by the participants were successful: they accelerated the learning process of the base RL algorithm. The same experiment was performed again, this time using the game Super Mario Bros. \cite{Karakovskiy2012} as a testbed. In the second experiment, the RL’s performance worsened when using the FAs. This evidence suggests that designing an effective FA is more challenging than simply using an RS technique in complex environments \cite{Rosenfeld2018}.

\textit{Hierarchical Decomposition} (HRL) is an effective approach to tackle high-dimensional state-action spaces by decomposing them into smaller sub-problems or temporarily extended actions \cite{Sutton1999, Dietterich2000, Lee2011, Usunier2016}. In an HRL setting, an expert can define a hierarchy of sub-problems that can be reused as skills in different applications. Although HRL has been successfully tested in different research areas \cite{Arzate2018, Bai2015, Lee2011}, there are no studies from the user-experience perspective in an interactive RL application.

\subsection{Types of Feedback} 
\label{sec::TypeFeedback}
In this section, we present the most common types of human feedback found in the literature. In particular, we will describe how each type of feedback delivers the user’s expertise to the different human knowledge integration methods that we explained in the previous section.

\subsubsection{Binary critique}
\label{subsec::critique}
The use of \textit{binary critique} to evaluate an RL model’s policy refers to binary feedback (positive or negative) that indicates if the last chosen action by the agent was satisfactory. This signal of human feedback was initially the only source of reward used \cite{Isbell2001}. This type of feedback was shown to be less than optimal because people provide an unpredictable signal and stop providing critiques once the agent learns the task \cite{Isbell2006}. 

One task for the RL designer is to determine whether a type of reward will be effective in a given application. For example, it was shown that using binary critique as policy information is more efficient than using a reward signal \cite{Knox2010, Thomaz2008}. Similarly, Griffith et al. \cite{Griffith13} propose an algorithm that incorporates the binary critique signal as policy information. From a user experience perspective, it has been shown that using critique to shape policy is unfavorable \cite{Krening2018}.

It is also worth noting that learning from binary critique is popular because it is an easy and versatile method of using non-expert feedback; the user is required to click only ``$+$'' and ``$-$'' buttons.

\textit{Heuristic Function} — The \textit{heuristic function} approach \cite{Bianchi2013, Martins2013} is another example of the critique feedback category. Instead of receiving binary feedback directly from the user, in this approach, the critique signal comes from a hand-engineered function. This function encodes heuristics that map state-action pairs to positive or negative critiques. The aim of the heuristic function method is to reduce the amount of human feedback. Empirical evidence suggests that this type of feedback can accelerate RL algorithms; however, more research is needed to test its viability in complex environments such as videogames \cite{Rosenfeld2018}.

\textit{Query} — The authors of the \textit{active inverse reward design} approach \cite{Mindermann2018} present a query-based procedure for inverse reward design \cite{Hadfield2017}. A \textit{query} is presented to the user with a set of sub-rewards, and the user then has to choose the best among the set. The sub-rewards are constructed to include as much information as possible about unknown rewards in the environment, and the set of sub-rewards is selected to maximize the understanding of different suboptimal rewards.       

\subsubsection{Scalar-valued critique}
In a manner similar to binary critique, with the \textit{scalar-valued} critique type of feedback, users evaluate the performance of a policy. In this case, the magnitude of the scalar-valued feedback determines how good or bad a policy is. This method can be used on its own to learn a reward function of the environment purely through human feedback. However, it has been shown that humans usually provide nonoptimal numerical reward signals with this approach \cite{Leike2018}.   

\subsubsection{Action Advice}
\label{subsec::actionAdvice}
In the \textit{action advice} type of feedback, the human user provides the agent with the action they believe is optimal at a given state, the agent executes the advised action, and the base RL algorithm continues as usual. From the standpoint of user experience, the immediate effect of the user’s advice on the agent’s policy makes this feedback procedure less frustrating \cite{Krening2018, Krening2019}. 

There are other ways to provide action advice to the agent, such as learning from demonstration \cite{Taylor2011}, inverse RL \cite{Ng1999, Ziebart2008, Ziebart2009}, apprentice learning \cite{Abbeel2004}, and imitation learning \cite{Christiano2017}. All these techniques share the characteristic that the base RL algorithm receives as input an expert demonstration of the task the agent is supposed to learn, which is ideal, as people enjoy demonstrating how agents should behave \cite{Amershi2014, Thomaz2008, Kaochar2011}.

\subsubsection{Guidance}
The last type of feedback, \textit{guidance}, is based on the premise that humans find it more natural to describe goals in an environment by specifying the object(s) of interest at a given time-step \cite{Thomaz2006, Suay2011}. This human knowledge leverages the base RL algorithm because the focus is on executing actions that might lead it to a goal specified by the user, which means that the RL algorithm also has access to the transition function of the dynamics in the environment.

\section{Recent Research Results} 
\label{sec::StateOfTheArt}
In this section, we survey the interactive RL research area for studies using each of the human knowledge integration methods. These methods define how user feedback is mapped onto the design dimensions of RL-based agents and thus play a critical role in interactive RL. Given the vast amount of work in this research area, we focus on the methods we believe to be the most influential.

\subsection{Reward Shaping}
The reward shaping (RS) method aims to mold the behavior of a learning agent by modifying its reward function to encourage the behavior the RL designer wants.

Thomaz and Breazeal analyzed the teaching style of non-expert users depending on the reward channel they had at their disposal in \cite{Thomaz2006}. Users were able to use two types of RS: positive numerical reward and negative numerical reward. These types of feedback directly modify the value function of the RL model. However, when the user gives negative feedback, the agent tries to undo the last action it performed; this reveals the learning progress to the user and motivates them to use more negative feedback, which achieves good performance with less feedback \cite{Thomaz2006}. 

Another finding in \cite{Thomaz2006} is that some users give anticipatory feedback to the bot; that is, users assume that their feedback is meant to direct the bot in future states. This analysis displays the importance of studying users' teaching strategies in interactive RL. We need to better understand the user’s preferences to teach agents, as well as how agents should provide better feedback about their learning mechanism to foster trust and improve the quality and quantity of users' feedback.  

Another RS strategy is to manually create \textit{heuristic functions} \cite{Rosenfeld2018, Bianchi2013} that incentivize the agent to perform particular actions in certain states of the environment. This way, the agent automatically receives feedback from the hand-engineered heuristic function. The type of feedback is defined by the RL designer, and it can be given using any of the feedback types reviewed in this paper (i.e., critique or scalar-value). The experiments conducted in \cite{Rosenfeld2018} demonstrate that using heuristic functions as input for an interactive RL algorithm can be a natural approach to injecting human knowledge in an RL method.

The main shortcoming of heuristic functions is that they are difficult to build and require programming skills. Although it has been investigated how non-experts build ML models in real life \cite{iMLYang2018}, there are not many studies on the use of more natural modes of communication to empower non-experts in ML to build effective heuristic functions that generalize well.     

The Evaluative Reinforcement (TAMER) algorithm uses traces of demonstrations as input to build a model of the user that is later used to automatically guide the RL algorithm \cite{Knox2009}.  

Warnell et al. later introduced a version of the TAMER algorithm \cite{Knox2009} modified to work with a deep RL model \cite{Warnell2018}. In addition to the changes needed to make the TAMER algorithm work with a function approximation that uses a deep convolutional neural network, the authors of \cite{Warnell2018} propose a different approach for handling user feedback. Instead of using a loss function that considers a window of samples, they minimize a weighted difference of user rewards for each individual state-action pair.

Similarly, the authors of \cite{Arakawa2018} propose an algorithm called DQN-TAMER that combines the TAMER and Deep TAMER algorithms. This novel combination of techniques aims to improve the performance of the learning agent using both environments and human binary feedback to shape the reward function of the model. Furthermore, Arakawa et al. experimented in a maze-like environment with a robot that receives implicit feedback; in this scenario, the RS method was driven by the facial expression of the user. Since human feedback can be imprecise and intermittent, mechanisms were developed to handle these problems.

Deep versions of interactive RL methods benefit mostly from function approximation, as use of this technique minimizes the feedback needed to get good results. This advantage is due to the generalization of user feedback among all similar states -- human knowledge is injected into multiple similar states instead of only one.

\subsection{Policy Shaping}
The policy shaping (PS) approach consists of directly molding the policy of a learning agent to fit its behavior to what the RL designer envisions.

A PS approach directly infers the user's policy from critique feedback \cite{Griffith2013, Loftin2016}. Griffith et al. introduced a Bayesian approach that computes the optimal policy from human feedback, taking as input the critiques for each state-action pair \cite{Griffith2013}. The results of this approach are promising, as it outperforms other methods, such as RS. However, PS experiments were carried out using a simulated oracle instead of human users. Further experiments with human users should be conducted to validate the performance of this interactive RL method from a user-experience perspective.

Krening and Feigh conducted experiments to determine which type of feedback — critique or action advice — creates a better user experience in an interactive RL setting \cite{Krening2018}. Specifically, they compared the critique approach in \cite{Griffith2013} to the proposed Newtonian action advice in \cite{Krening2018_2}. Compared to the critique approach, the action advice type of feedback got better overall results: it required less training time, it performed objectively better, and it produced a better user experience with it.

MacGlashan et al. introduced the Convergent Actor-Critic by Humans (COACH) interactive RL algorithm \cite{Macglashan2017}. There is also an extension to this work named deep COACH \cite{Arumugam2018} that uses a deep neural network coupled with a replay memory buffer and an autoencoder. Unlike the COACH implementation in \cite{Macglashan2017}, deep COACH \cite{Arumugam2018} uses raw pixels from the testbed as input. The authors argue that using this high-level representation as input means their implementation is better suited for real scenarios. However, the testbed consists of simplistic toy problems, and a recent effort demonstrated that deep neural networks using raw pixels as input spend most of their learning capacity extracting useful information from the scene and just a tiny part on the actual behaviors \cite{Cuccu2019}.

\subsection{Guided Exploration Process}
Guided exploration process methods aim to minimize the learning procedure by injecting human knowledge to guide the agent’s exploration to states with a high reward.

Thomaz and Breazeal conducted human-subject experiments in the Sophie's Kitchen platform to evaluate the performance of diverse human knowledge integration methods \cite{Thomaz2006}. Their results suggest that using guidance feedback is more effective than using scalar reward feedback. Based upon the algorithm in \cite{Thomaz2006}, Suay et al. proposed a variation in which the user can guide exploration by pointing out goal states in the environment \cite{Suay2011}; the bot then biases its policy to choose actions that lead to the indicated goal. These experiments were generally successful, but their highlight is their finding that using only exploration guides from the user produces the best results and reduces the amount of user feedback needed \cite{Suay2011}.


 
There have been efforts to create adaptive shaping algorithms that learn to choose the best feedback channels based on the user's preferred strategies for a given problem \cite{Yu2018}. For instance, Yu et al. defined an adaptive algorithm with four different feedback methods at its disposal that are based on exploration bias and reward shaping techniques \cite{Yu2018}. To measure the effectiveness of the feedback methods at every time-step, the adaptive algorithm asks for human feedback; with this information, a similarity measure between the policy of the shaping algorithms and the user's policy is computed. Then, according to the similarity metric, the best method is selected using a softmax function, and the value function for the selected method is updated using q-learning. Once the adaptive algorithm has enough samples, it considers the cumulative rewards to determine which feedback methodology is the best. Overall, one of the exploration bias-based algorithms produced better results and was chosen most often by the adaptive algorithm, as well as demonstrating good performance on its own. The authors of \cite{Yu2018} call this algorithm action biasing, which uses user feedback to guide the exploration process of the agent. The human feedback is incorporated into the RL model using the sum of the agent and the user value functions as value functions. 

In general, using human feedback as guidance for interactive RL algorithms appears to be the most effective in terms of performance and user experience. However, to make human feedback work, it is necessary to have a model of the environment, so the interactive RL algorithm knows which actions to perform to reach the state proposed by the user. This can be a strong limitation in complex environments where precise models are difficult to create.

\subsection{Augmented Value Function}
The procedure to augment a value function consists of combining the value function of the agent with one created from human feedback.

Studies have proposed combining the human and agent value functions to accelerate learning \cite{Knox2010, Taylor2011}. In \cite{Taylor2011}, the authors introduce the Human-Agent Transfer (HAT) algorithm, which is based on the rule transfer method \cite{Taylor2007}. The HAT algorithm generates a policy from the recorded human traces. This policy is then used to shape the q-value function of the agent. This shaping procedure gives a constant bonus to state-action pairs of the agent q-learning function that aligns with the action proposed by the previously computed human policy.

Brys et al. present an interactive RL algorithm named $\text{RLfD}^2$ that uses demonstrations by an expert as input \cite{Brys2015}. With these demonstrations, they create a potential-based piecewise Gaussian function. This function has high values in state-action pairs that have been demonstrated by the expert and $0$ values where no demonstrations were given. This function is used to bias the exploration process of a Q($\lambda$)-learning algorithm in two different ways. First, the Q-function of the RL algorithm is initialized with the potential-based function values. Second, the potential-based function is treated as a shaping function that complements the reward function from the environment. The combination of these two bias mechanisms is meant to leverage human knowledge from the expert throughout the entire learning process.

From the user-experience standpoint, the augmented value function design dimension has the advantage of \textit{transfer learning}. For instance, a value function constructed by one user can be used as a baseline to bias the model of another user trying to solve the same task — the learned knowledge from one user is transferred to another. Multiple sources of feedback (coded as value functions) can be combined to obtain more complete feedback in a wider range of states. It is also convenient that a model of the environment is not essential for this approach.

\subsection{Inverse Reward Design}
\textit{Inverse reward design} (IRD) is the process of inferring a true reward function from a proxy reward function.

IRD \cite{Hadfield2017, Mindermann2018} is used to reduce reward hacking failures. According to the proposed terminology in \cite{Hadfield2017}, the hand-engineered reward function named the \textit{proxy reward function} is just an approximation of the true reward function, which is one that perfectly models real-world scenarios. The process of inferring a true reward function from a proxy reward function is IRD. 

To infer the true reward function, the IRD method takes as input a proxy reward function, the model of the test environment, and the behavior of the RL designer who created it. Then, using Bayesian approaches, a distribution function that maps the proxy reward function to the true reward function is inferred. This distribution of the true reward function makes the agent aware of uncertainty when approaching previously unseen states, so it behaves in a risk-averse manner in new scenarios. The results of the experiments in \cite{Hadfield2017} reveal that reward hacking problems lessen with an IRD approach.

The main interaction procedure of IRD starts as regular RS, and the system then queries the user to provide more information about their preference for states with high uncertainty. This kind of procedure can benefit from interaction techniques to better explain uncertainty to humans and from compelling techniques to debug and fix problems in the model.

\section{Design Principles for interactive RL}
\label{sec::DesignPrinciples}
The design of RL techniques with a human-in-the-loop requires consideration of how human factors impact the design of and interaction with these machine learning algorithms. To produce behaviors that align with the user's intention, clear communication between the human and the RL algorithm is key. In this section, we will introduce general design principles for interactive RL meant to guide HCI researchers to create capable and economical interactive RL applications that users can understand and trust.

\subsection{Feedback}
\textit{Delay of human feedback} has a great impact on the performance of interactive RL applications. Delay refers to the time that a human user needs to evaluate and deliver their feedback to the interactive RL algorithm. Many examples propose different strategies to deal with feedback delay \cite{Arakawa2018, Arumugam2018, Isbell2001, Thomaz2005, Thomaz2006, Warnell2018, Peng2016}. 

The most common approach consists of a delay parameter that expresses to how many past time-steps the current feedback will be applied; this parameter is usually found in practice \cite{Arumugam2018}. Warnell et al. conducted an experiment to quantify the effect of different reward distribution delays on the effectiveness of their algorithm Deep TAMER \cite{Warnell2018}. This experiment revealed that a small change in the parameters of the expected user feedback timing distribution can have a great impact on performance; expert knowledge on this timing distribution is key to achieving good results.

A different approach to deal with feedback delay is proposed in \cite{Suay2011}. In this case, the interactive RL algorithm pauses the agent exploration process every time-step to give the user time to give feedback. More elaborated approaches assume that the feedback delay follows a probability distribution \cite{Arakawa2018}. Likewise, it has been found that less avid users need more time to think and decide on their feedback \cite{Peng2016}. It is therefore important to adapt the feedback delay depending on users’ level of knowledge. 

\textit{Fatigue} of users and its effects on the quantity and quality of feedback should be considered. It has been observed that humans tend to reduce the quantity of feedback they give over time \cite{Isbell2001, Ho2015, Knox2012_3}. The quality also diminishes, as humans tend to give less positive feedback over time. According to \cite{Macglashan2017}, this degradation of the quantity and quality of human feedback also depends on the behavior exhibited by the agent. The authors of \cite{Macglashan2017} found that humans tend to give more positive feedback when they notice that the agent is improving its policy over time: feedback is therefore policy-dependent \cite{Miltenberger2011}. On the other hand, the experiments of \cite{Christiano2017} offer evidence to support that human users gradually diminish their positive feedback when the agent shows that it is adopting the proposed strategies. Fachantidis et al. performed experiments to determine the impact of the quality and distribution of feedback on the performance of their interactive RL algorithm \cite{Fachantidis2017}.

\textit{Motivating users to give feedback} — elements of gamification have been adopted with good results; gamification strategies have been shown to improve the quantity and quality of human feedback \cite{Lessel2019, Li2018_2}.

Some studies focus on improving the quality and quantity of human feedback by incorporating an active question procedure in the learning agent \cite{Griffith2013, Li2013, Amir2016, Fachantidis2017}; that is, the agent can ask the user to give feedback in particular states. Amir et al. present an active interactive RL algorithm in which both the agent and the demonstrator (another agent) work together to decide when to make use of feedback in a setting with a limited advice budget. First, the agent determines if it needs help in a particular state and asks the demonstrator for attention. Depending on the situation of the agent, the demonstrator determines if it will help or not. The results of these experiments are promising because they achieve a good level of performance while requiring less attention from the demonstrator. 


\textit{Maximizing the use of feedback} is necessary because interactive RL in complex environments might require up to millions of interaction cycles to get good results \cite{Christiano2017, Mnih2015}. It has been demonstrated that the inclusion of an approximation function that propagates human feedback to all similar states is effective to tackle the sample inefficiency of interactive RL in complex environments such as videogames \cite{Warnell2018}.

\subsection{Typification of the End-user}
We found that the most important features to typify an end-user in interactive RL involves:

\textit{Knowledge level in the task at hand} — this has an impact on the quality, quantity, and delay of feedback, and could limit the use of some feedback types that require more precise information, such as demonstrations or action advice.

\textit{Preferred teaching strategy} — this is essential to select the feedback type appropriate for a given application (see Figure \ref{Fig:iRL}).

\textit{Safety concerns} — this refers to the severity of the effects of the agent’s actions if an error in its behavior occurs. If the severity of the errors is high (e.g. property damage, personal harm, politically incorrect behavior, etc.), end-users’ perception and willingness to interact with the agent will diminish \cite{Morales2019}. The quality and quantity of feedback are therefore also affected. 

An end-user typification using these features would help researchers to select the best combination of feedback type and design dimension for a particular interactive RL and for the expected type of end-user. For instance, although all design dimensions assume that the user knows a policy that is at least good enough to solve the task, some design dimensions can better handle non-optimal policies from human feedback \cite{Griffith2013}. Empirical evidence suggests that the exploration process is the most economical design dimension for interactive RL applications \cite{Thomaz2006, Griffith2013, Krening2018}. Nonetheless, myopic agents — which use the policy design dimension — have demonstrated great success in some tasks \cite{Fachantidis2017, Leike2018}. 

The hierarchical decomposition and function approximation design dimensions require human users with a deep understanding of the problem to make effective use of them. Another possibility is combining different design dimensions in the same interactive RL method. It would also be useful to design an active interactive RL that learns which design dimension best suits the application and type of use; in this way, the interactive RL algorithm can minimize the feedback needed from the user \cite{Griffith2013}. For example, a combination of function approximation and policy design dimensions would enable the use in a videogame environment of an interactive RL that needs only $15$ minutes of human feedback to get positive results \cite{Warnell2018}.

\subsection{Fast Interaction Cycles}
In an iterative procedure, such as an interactive RL algorithm, a fast evaluation of the agent’s behavior can substantially reduce the time needed for each teaching loop. Some of the approaches meant to lessen the evaluation process consist of evaluation, visualization, or explanatory techniques.

\textit{Evaluation techniques} such as  queries, in which the user selects the best reward function among a set \cite{Hadfield2017, Sorensen2016}, the presentation of multiple trajectories by the agent that then summarizes its behavior \cite{Wilson2012}, or crowd evaluations that distribute the evaluation task among various users \cite{Lelis2015}.

\textit{Visualization techniques}, such as the HIGHLIGHTS algorithm \cite{Amir2018}, focus on creating summaries of agents so people can better understand an agent's behavior in crucial situations. These summaries are sets of trajectories that provide an overview of the agent's behavior in key states. It is therefore fundamental to define a metric that measures the importance of states. This is calculated as the difference between the highest and lowest expected returns for a given state. That is, a state in which taking a particular action leads to a significant decrease in the expected reward is considered important.

\textit{Explanatory techniques} are a different way to address the evaluation problem by explaining the agent’s decision process to the human user. This approach enhances the user's understanding of how the agent learns, interprets, and determines the best action to take in a given state.  

One example of uncertainty explanation is introduced in \cite{Thomaz2006}. The authors use the gaze of the agent in their experiments as an indicator of its level of uncertainty in any given state. 

The main idea of the gaze mechanism is to make the agent explicitly express at every time-step the options it considers most promising by staring at the object of interest. If there is more than one promising action in a given state, the agent will have to focus its gaze on different objects in the environment, communicating indecision to the user and giving the user time to give feedback to the agent. The authors also found that users tend to give more feedback when the agent displays uncertainty, which optimizes the exploration process of the robot.



\subsection{Design Implications}
Our analysis has found some design factors that need more exploration to better understand their impact on interactive RL. These include adaptive systems that can choose between different design dimensions, enabling the use demonstration as feedback in complex robot environments where it is difficult for the user to show the agent how to perform the task it has to learn. This makes interactive RL based applications accessible to more types of users (e.g. non-experts in RL, non-experts in the task at hand, those with physical limitations, etc.), so this design dimension is better for non-experts in the task at hand.

\section{Open Challenges}
\label{sec::OpenChallenges}
Below, we list what we consider the most promising research directions in interactive RL. 

\subsection{High-dimensional Environments}
Making interactive RL feasible for real-world applications with high-dimensional environments is still a challenge. Broadening the domains and applications in which interactive RL is useful is key to extending the impact of ML in people's life. 

Recent interactive RL applications that use an autoencoder to spread human feedback to similar states perform well in high-dimensional environments \cite{Arumugam2018}. One underexploited approach is using crowd-sourced methods to provide human feedback \cite{Isbell2001} or build a reward function.

\subsection{Lack of Evaluation Techniques}
The evaluation of interactive RL applications from the HCI perspective is limited; most studies focus on improving the performance of algorithms rather than the interaction experience, but proper evaluation is essential to create practical applications for different types of end-users. For instance, the studies by Krening et al. \cite{Krening2018, Krening2019} present an adequate evaluation of the user experience in an interactive RL setting. 

In the same context, the \textit{problem of generalization from simple testbeds} (e.g., GridWorld) exists, as the results are not consistent when using more complex testbeds. For example, even though RS has been demonstrated to be effective and easy to use by all types of users in a GridWorld environment \cite{Arakawa2018, Arumugam2018}, using the same human knowledge integration method to train agents for the Infinite Mario platform is challenging and ineffective, even for expert users \cite{Rosenfeld2018}.

\subsection{Lack of Human-like Oracles}
The use of simulated oracles in the early stages of implementation of interactive RL applications is useful to save time and get an idea of the results with human users \cite{Cruz2015, Yu2018, Griffith2013}. However, it has been found that results with simulated oracles can differ from those with human users \cite{Krening2018, Yu2018} — simulated oracles do not behave like humans. More studies are needed to determine what features make human feedback different from simulated oracles.

\subsection{Modeling Users}
As far as we know, a formal model of users for interactive RL (or IML in general) has not been proposed. Such a model could be used to create interactive RL applications with an active initiative that avoids user fatigue by detecting and predicting user behavior. That is, the interactive RL system would ask for help from the user with the correct frequency to avoid user fatigue. Another possibility is implementing RL-based applications that adapt the interaction channel and feedback type according to the user’s preferences. This would require empirical studies to find a way to map between user types and their preferred interaction channels (see next subsection).

\subsection{Combining Different Design Dimensions}
A better understanding of the strengths and weaknesses of each design dimension and feedback type in interactive RL would lead the community to develop effective combinations of interactive RL approaches. Achieving this would require extensive user experience studies with all different combinations of design dimensions and feedback types, as well as in testbeds with small and big state spaces. Furthermore, it would be favorable to find a mapping that considers the type of end-user. This, in combination with an end-user model, would enable the design of interactive RL applications that adapt to the current end-user.

\subsection{Safe interactive RL}
With a human-in-the-loop that can teach an agent new skills, it is important to have a mechanism to restrict agents from going to negative states or performing actions that lead to dangerous situations for humans, politically incorrect situations, or unwanted bias. The addition of Safe RL techniques \cite{Garcia2015} to an interactive RL setting can be an excellent starting point. Safe RL techniques ensure good performance and respect safety constraints during learning and deployment. Similarly, exploring new interaction methods to impose limits on interactive RL models is a promising research direction.

\subsection{Fast Evaluation of Behaviors}
Fast evaluation methods through visualization techniques have not been adequately studied. These approaches include applications to summarize behavior \cite{Amir2018, Hadfield2017, Sorensen2016, Wilson2012} and explain uncertainty \cite{Schneider2019, Li2016}. The main goal of these methods is to reduce the time needed to evaluate the behavior of agents, and assist users to give higher quality feedback.

\subsection{Explainable interactive RL}
RL-based agents have a particular way of representing their environment, depending on the sensors at their disposal and how they interpret the incoming signals. It can therefore be complicated for humans to elicit the state representation of agents, which can lead to giving incorrect feedback to agents. Explainable interactive RL methods aim to help people understand how agents see the environment and their reasoning procedure for computing policies \cite{Liu2019, Krening2019, Wang2019, Adadi2018}. With more transparency in an agent model, people could find better ways to teach agents.

Another way to take advantage of a transparent agent is by debugging its model. That is, users would be able to interpret why the agent made a certain decision in a particular state. This feature would help people find errors or unwanted bias in the model \cite{Myers2020}, which is essential in applications that could have a substantial impact on people's lives \cite{EUdataregulations2018}.


\section{Conclusions}
We have presented a survey of interactive RL to empower HCI researchers with the technical background in RL needed to design new interaction techniques and applications. We strongly believe that our paper is a step toward the wider use of interactive RL in the HCI community.

\section{Acknowledgements}
This work was supported by JST CREST Grant Number JPMJCR17A1, Japan. 

\balance{}

\bibliographystyle{SIGCHI-Reference-Format}
\bibliography{sample}

\end{document}